\documentclass[journal=jacsat,manuscript=article]{achemso}

\usepackage[version=3]{mhchem} 
\usepackage[T1]{fontenc}  
\usepackage{subfigure}     

\author{Xu-Chen Wang}
\affiliation[Aalto University]{Department of Electronics and Nanoengineering, Aalto University, Espoo, Finland}

\email{xuchen.wang@aalto.fi}
\phone{+358-503097794}

\author{Sergei A. Tretyakov}
\affiliation[Aalto University]{Department of Electronics and Nanoengineering, Aalto University, Espoo, Finland}

\title[An \textsf{achemso} demo]
  {Graphene-Based Tunable Metasurface for All-angle Perfect Absorption}

\abbreviations{IR,NMR,UV}

\begin{document}

\begin{abstract}
 Motivated by the idea of ``smart'' metasurfaces, we will demonstrate a graphene-based tunable absorber in which perfect absorption can be achieved for all angles of incidence,  only by tuning the Fermi level of graphene. We place an unpatterned graphene sheet on a mushroom-type high impedance surface whose resonant frequency is stable for all incident angles. For TM-polarization, perfect absorption can be realized from normal to grazing incidence at the same frequency when modulating the Fermi level of graphene from 0.18~eV to 1~eV.
\end{abstract}

\section{Introduction}
Future developments of ``smart'' metasurfaces which change the functionality using programmable/tunable platforms will provide adaptive response. In this new paradigm, the development of tunable absorbers capable to adapt the performance to different operation frequencies, rates of absorption or illumination angles are necessary.  Due to the tunable conductivity, graphene has excited great interest for this mission. 
Recently, a systematic approach for realizing tunable perfect absorption in low-quality graphene has been presented in \cite{wang2017tunable}. 
Although large tunability of the absorption  frequency and the rate of absorption are achieved in that work, 
full absorption was still restricted for only normal incidence. 
Here, we propose and explore a possibility of tunable perfect absorption for all illumination angles.

It is well known that conventional thin  absorbers  can provide  perfect absorption only at one predefined incidence angle, $\theta_{\rm i}$. The reflectance increases with the incidence angle  deviating from this specified angle due to the impedance mismatch between the incident wave and the absorbing structure at other incidence angles.
For example, consider a Salisbury screen, in which a resistive sheet is positioned on top of a grounded quarter-wavelength substrate acting as a high impedance surface (HIS).
If the sheet impedance of the resistive film is matched to the wave impedance ($Z_{\rm w}=120\pi\cos\theta_{\rm i}$ for TM polarization) no energy will be reflected back. When the incidence angle changes, the wave impedance changes, and the sheet resistance is not any longer matched. Moreover, the electrical thickness of the substrate changes, and it does not act as a HIS at the operational  frequency. 
These two factors lead to the degradation of  absorption. 
Many efforts have been made to improve the angular stability of absorbers, for example, making the resonance of HIS insensitive to the incidence angles \cite{zhirihin2017mushroom}. 
However, even with this solution, the angular spectrum of absorption is still limited since the sheet resistance of the lossy sheet does not follow the change of the wave impedance.

\begin{figure}[h!]
	\centering
	\subfigure[]{
		\includegraphics[width=0.5\linewidth]{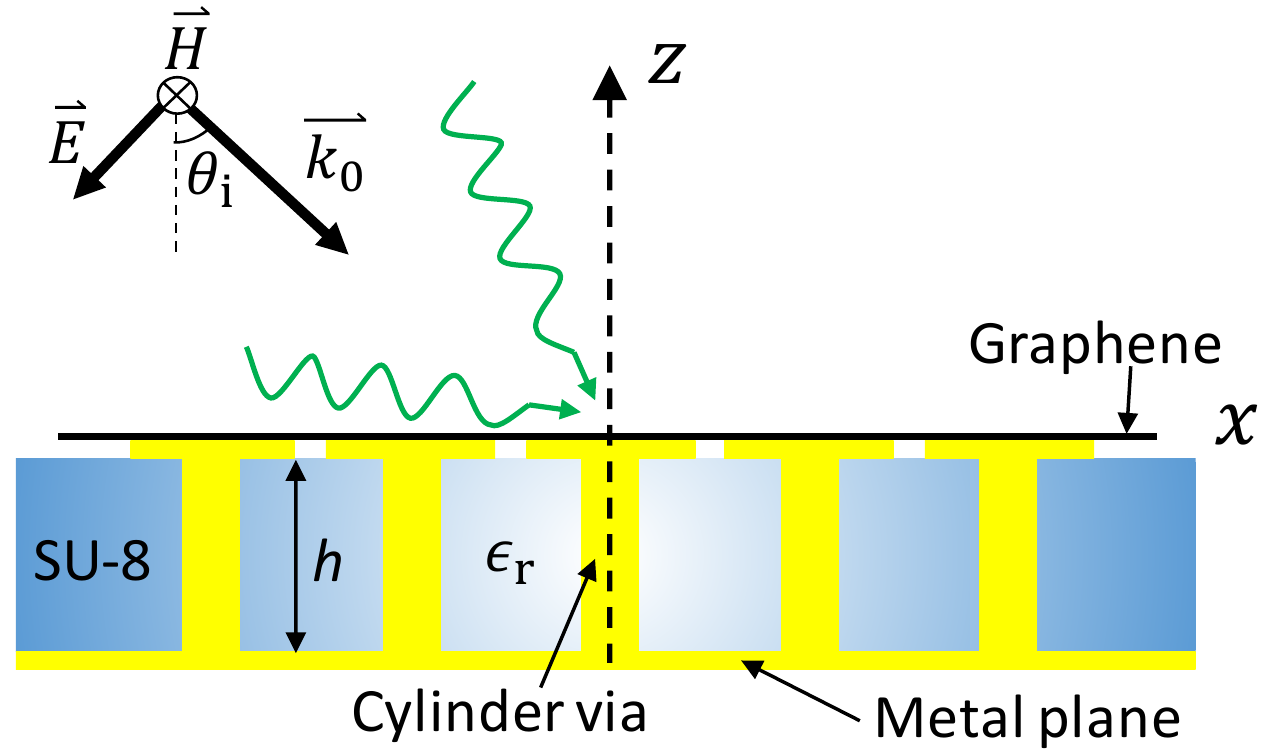}
		\label{fig:side_view}}
	\subfigure[]{
		\includegraphics[width=0.3\linewidth]{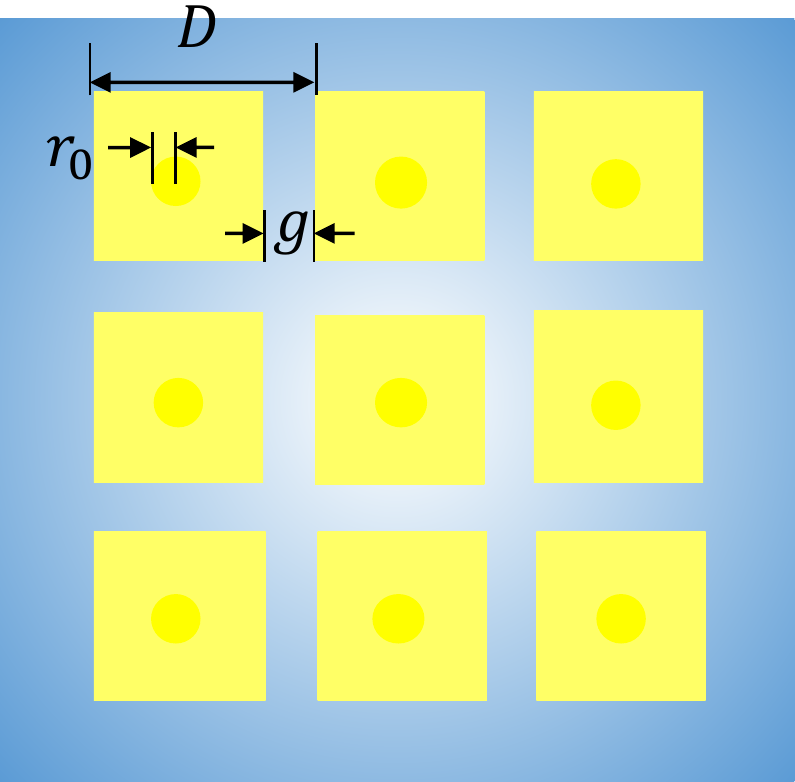}
		\label{fig:top_view}}
	\subfigure[]{
		\includegraphics[width=0.35\linewidth]{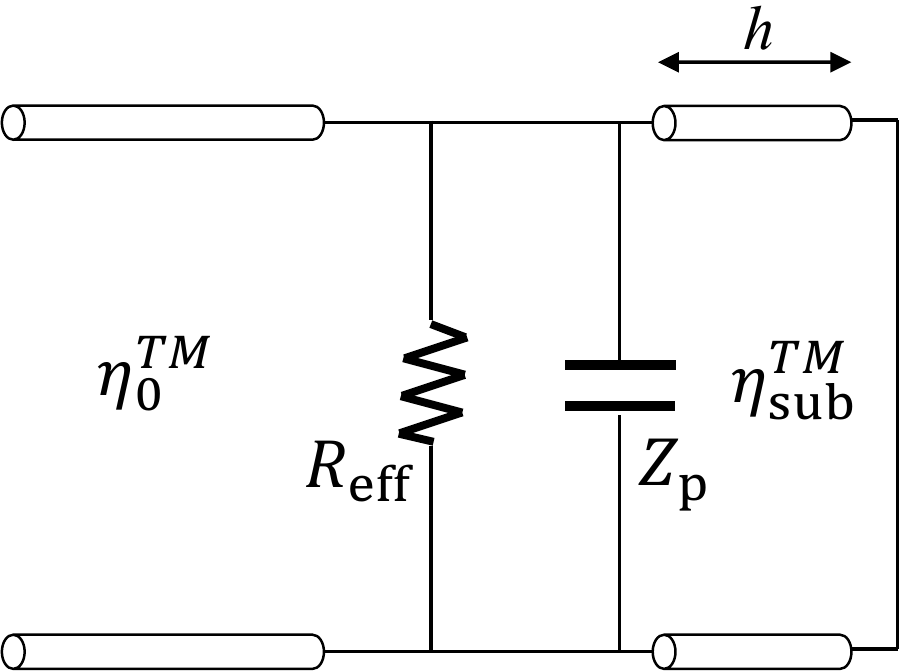}
		\label{fig:circuit}}
	\caption{Proposed structure for all-angle perfect absorbers: (a) representation of the structure and the incident wave,  (b) top view of the structure, and (c) equivalent circuit of the system.}\label{structure}
\end{figure}	
	
For overcoming the limitations of known designs, we propose the structure shown in Fig.~\ref{fig:side_view}.
We use a layer of unpatterned graphene as a resistive load and place it over a mushroom-type HIS. Based on the angularly stable resonance of the mushroom structure, complete absorption for TM-polarized wave can be realized at all incidence angles by tuning the Fermi level of graphene.

\section{Tunability of the graphene sheet impedance and angular-stability of the HIS}

Our design approach is based on the use of a HIS whose resonant frequency does not change with the incidence angle and a tunable resistive sheet that does not perturb the performance of the HIS. In what follows we will demonstrate that the structure shown in Fig.~\ref{fig:side_view} satisfies these conditions.

\begin{figure}[h!]
	\centering
	\subfigure[]{
		\includegraphics[width=0.5\linewidth]{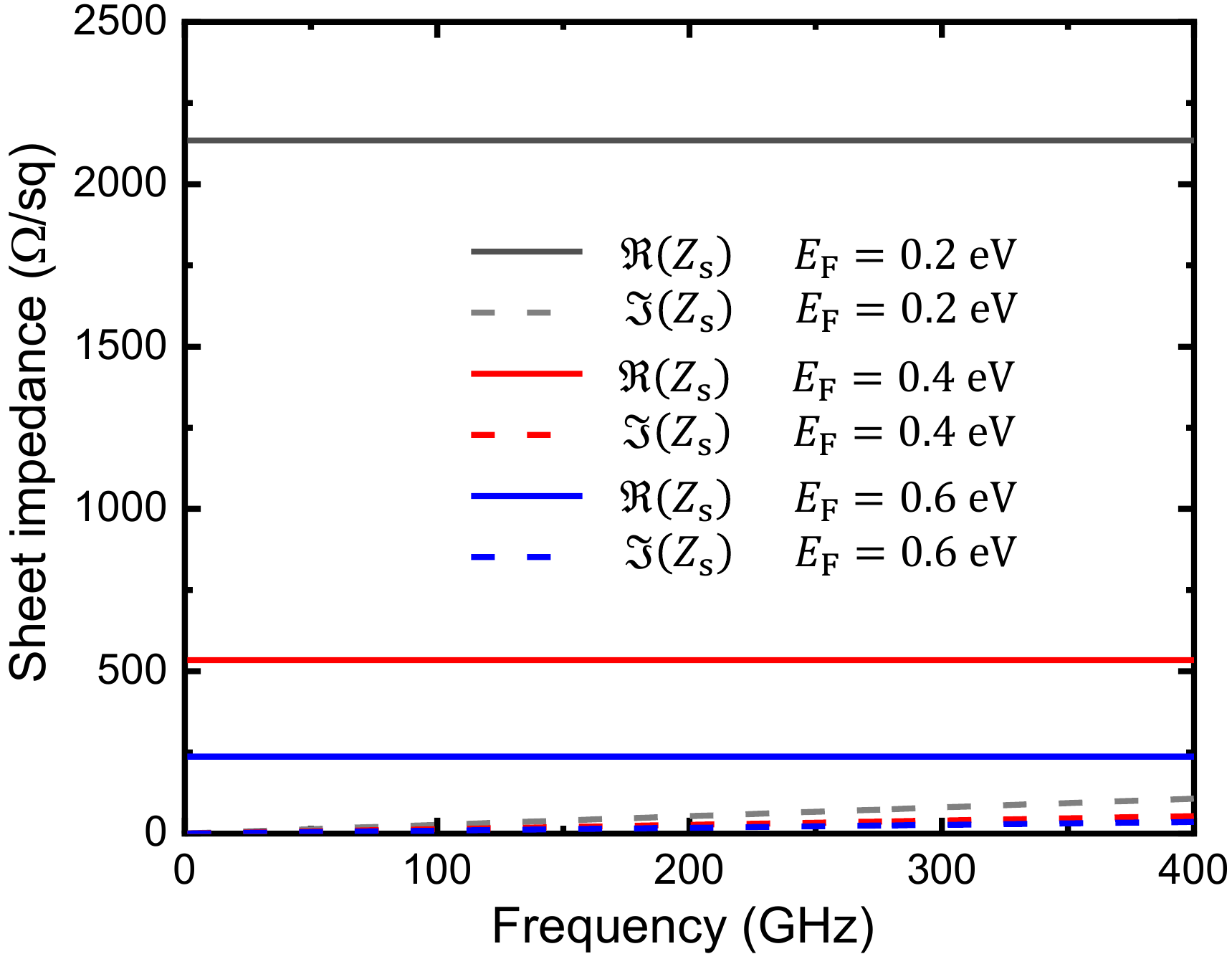}
		\label{fig:Sheet_impedance_graphene}}
	\subfigure[]{
		\includegraphics[width=0.5\linewidth]{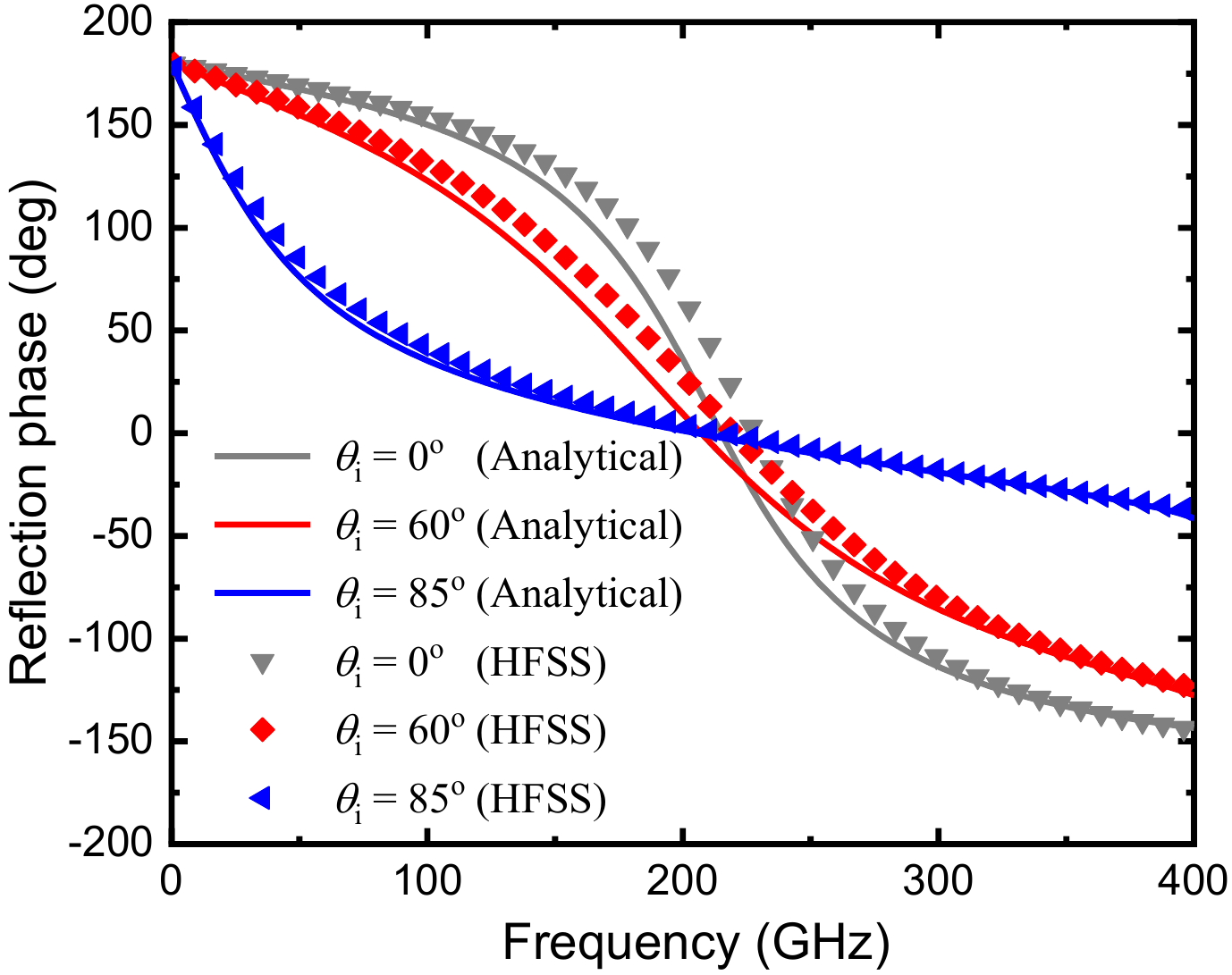}
		\label{fig:reflection phase}}
	\caption{Tunability of the sheet impedance and angular stability of the HIS: (a) Sheet impedance of graphene calculated from the Drude model. The carrier mobility is  $\mu=1000~{\rm cm}^{2}~{\rm V}^{-1}~{\rm s}^{-1}$; (b) Reflection phase of  the mushroom-type HIS. Here, $D=200~\mu$m, $g=25~\mu$m, $r_0=30~\mu$m and $\epsilon_{\rm r}=2.8(1-j0.045)$.}\label{sheet impedance and phase resoponse}
\end{figure}

In the low THz-band and below, where the interband transitions in graphene are very week, the surface conductivity of monolayer graphene can be modelled by the Drude formula, $\sigma_{\rm sg}=-je^2E_{\rm F}/{\pi\hbar^2(\omega-j\gamma)}$. Here, $e$ is the electron charge, $\hbar$ is the Plank constant, and $\gamma=e{v_{\rm F}}^2/\mu E_{\rm F}$ is the scattering rate of electrons which is related to the carrier mobility $\mu$ and the Fermi level of graphene $E_{\rm F}$.
The sheet impedance of graphene $Z_{\rm sg}=1/\sigma_{\rm sg}$ is a complex value denoted as $Z_{\rm sg}=R_{\rm sg}+jX_{\rm sg}$. 
If the frequency of the incident wave  $\omega$ is much smaller than the scattering rate of graphene (e.g. millimeter band or below), the inductive part of the sheet impedance can be neglected and the monolayer graphene behaves as a purely resistive sheet. 
Figure~\ref{fig:Sheet_impedance_graphene} shows the surface impedance of graphene from 0~GHz to 400~GHz for different values of the Fermi level.
Obviously, the sheet resistivity is frequency independent, and the sheet inductance is a linear function of the frequency but rather small compared to its resistive part. 
The series expression of sheet impedance can be equivalently transformed into its shunt form, $Z_{\rm sg}=(R_{\rm sg}+\frac{{X_{\rm sg}}^2}{R_{\rm sg}})\parallel j(X_{\rm sg}+\frac{{R_{\rm sg}}^2}{X_{\rm sg}})$. 
If $R_{\rm sg}\gg X_{\rm sg}$, the huge shunt reactance is equivalent to  an open circuit. Therefore, we can conclude that  graphene at low frequencies will not affect the resonant frequency of any resonant structure placed below it.

The schematics of the mushroom structure comprising an array of patches over a dielectric layer perforated with metallic vias is shown in Figs.~\ref{fig:side_view}~and~\ref{fig:top_view}. Since the metallic vias have no response to the transverse electrical field ($r_0\ll D$), the wire medium is viewed as an uniaxial material with effective permittivity
${\overline{\overline{\epsilon}}}_{\rm eff }=\epsilon_{\rm r}(\textbf{u}_x\textbf{u}_x+\textbf{u}_y\textbf{u}_y)+\epsilon_{zz}\textbf{u}_z\textbf{u}_y$. If the wavelength is much larger than the substrate thickness, the spatial dispersion in the wire medium  can be neglected. In this situation we can use the quasi-static approximation to describe the vertical permittivity (the local model), $\epsilon_{zz}=\epsilon_{\rm r}(1-{k_{\rm p}}^2/{k_{\rm r}}^2)$, where $k_{\rm p}$ is the plasma wave number related to the structural parameters of metallic wires \cite{tretyakov2003analytical}. For TM incidence, the surface impedance of grounded wire medium slab is inductive and given as $Z_{\rm sw}=j\omega\mu_0h[1-{k_0}^2{\sin^2\theta}/({k_{\rm r}}^2-{k_{\rm p}}^2)]$ \cite{tretyakov2003analytical}. For low frequencies we have $k_{\rm r}\ll k_{\rm p}$, and the effective inductance of $Z_{\rm sw}$ is insensitive to the incident angle. 
The total impedance of the HIS is a shunt connection of the grounded wire medium slab and the capacitive patch array, ${Z^{-1}_{\rm HIS}}=Z^{-1}_{\rm p}+Z^{-1}_{\rm sw}$. The effective capacitance of the patch array is also angularly stable for TM-polarized waves \cite{tretyakov2003analytical}. Since both of the effective capacitance and inductance are angle-insensitive, the resonant frequency of the HIS should also be independent of the incidence angle. 
In Fig.~\ref{fig:reflection phase}, we analytically and numerically calculate the reflection phase of the mushroom structure, and observe that the resonant frequency of HIS keeps practically unchanged from normal to grazing incidence.

\section{Graphene on mushroom-type high impedance surface}

Next, we study the absorption when a continuous  graphene layer is placed over a mushroom-type HIS. In order to achieve total absorption, the sheet impedance of graphene should be equal to the wave impedance. We note that the graphene sheet resistance (from several hundreds to several thousands) is much larger than the free space impedance at large incidence angles. However, in this structure the patch array can effectively reduce the sheet impedance of graphene, as demonstrated in the reference \cite{wang2017tunable}.  The effective shunt resistance can be expressed as $R_{\rm eff}=R_{\rm sg}/p$, where $p=(D-g)/g$ is the geometry factor. 
The total input impedance is calculated as $Z^{-1}_{\rm inp}=(R_{\rm eff})^{-1}+Z^{-1}_{\rm HIS}$, and the reflection coefficient reads $R=(Z_{\rm inp}-Z_{\rm w}/(Z_{\rm inp}+Z_{\rm w})$.

\begin{figure}[h!]
	\centering
	\subfigure[]{
		\includegraphics[width=0.5\linewidth]{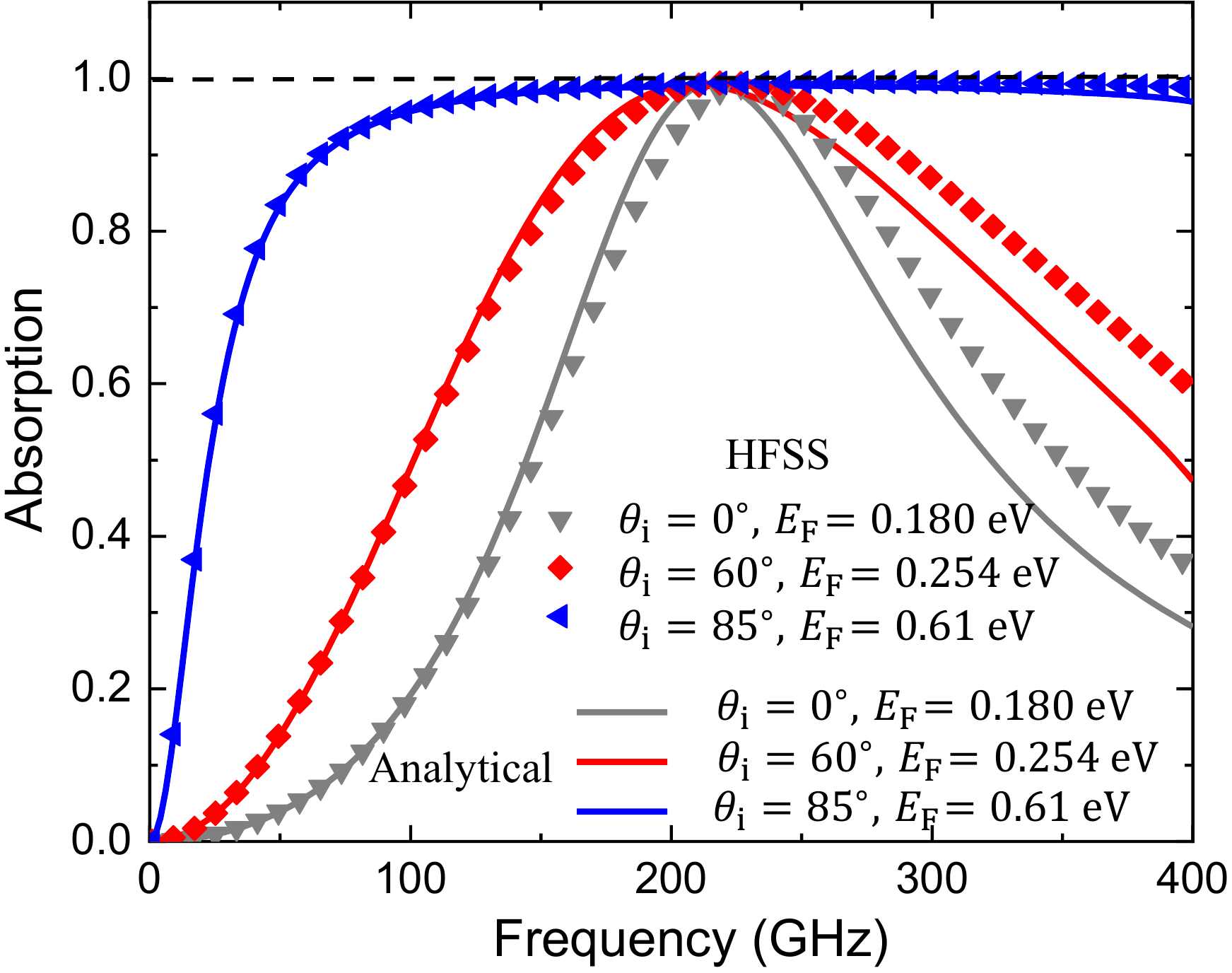}
		\label{fig:absorption}}
	\subfigure[]{
		\includegraphics[width=0.5\linewidth]{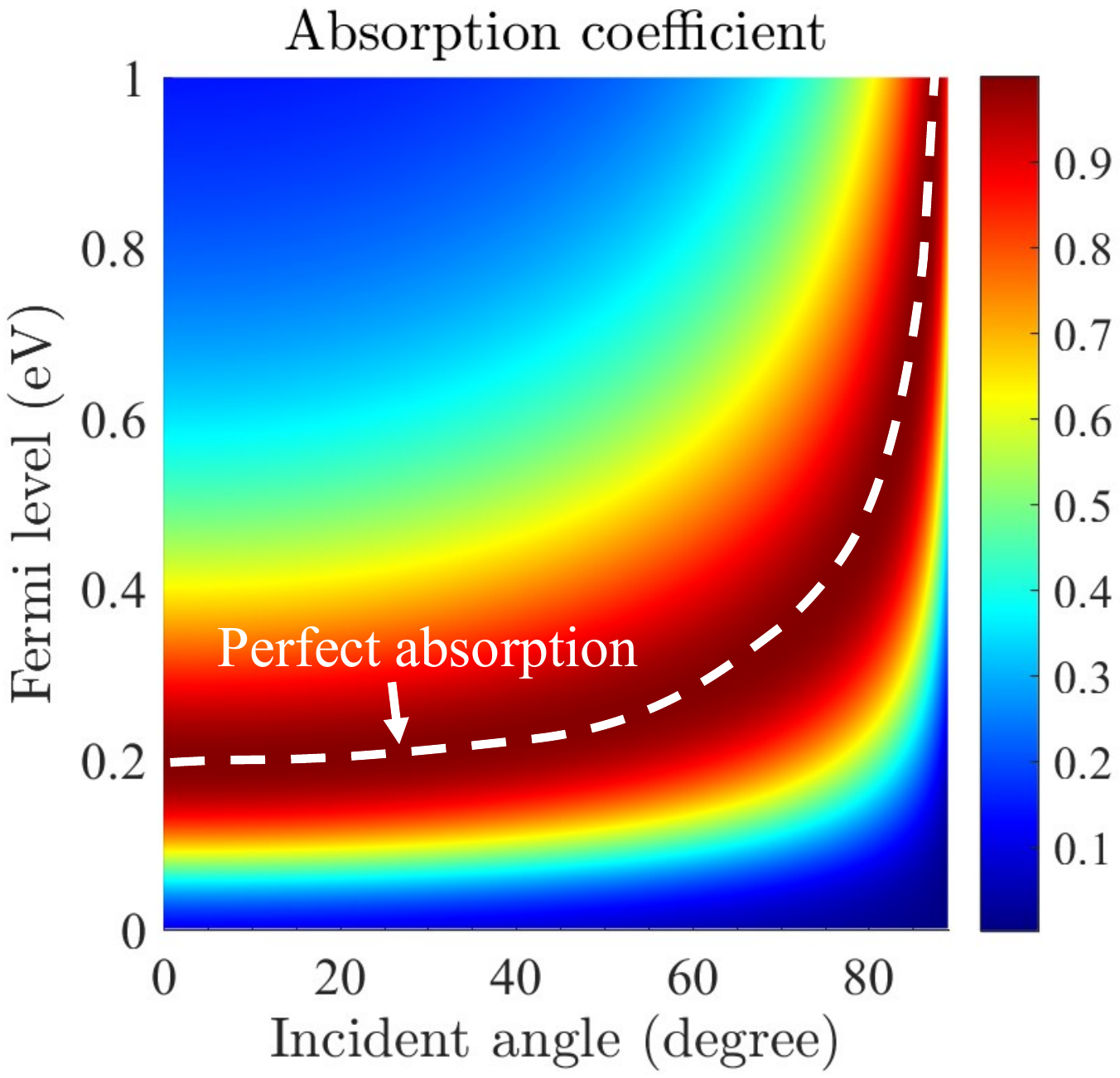}
		\label{fig:absorption_color}}
	\caption{Systematic control of the absorption at different angles: (a) Analytical and numerically calculated  absorption coefficients at different incidence angles; (b) Analytical and calculated absorption coefficients in terms of the incidence angles and the Fermi levels of graphene at 220~GHz.}\label{absorption}
\end{figure}

Figure~\ref{fig:absorption} displays the absorption coefficient ($A=1-|R|^2$) at different incidence angles with the appropriate Fermi levels of graphene, realizing perfect absorption at 220~GHz.
With the increase of the incidence angle, the bandwidth of the absorber is greatly broadened. This is due to the intrinsic broadband property of HIS at oblique TM-illumination. 
Figure~\ref{fig:absorption_color} shows the absorption coefficient with respect to graphene Fermi level as well as the incidence angle at the resonance of the HIS. From $\theta_{\rm i}=0^\circ$ to $\theta_{\rm i}=88^\circ$, we can always find a value of $E_{\rm F}$ to obtain perfect absorption at every angle. Interestingly, at each incidence angle the absorption level can be tuned in a wide range  only by modulating $E_{\rm F}$.

\null\vspace{-6mm}

\section{Conclusion}
In this work, we have demonstrated an  angle-insensitive tunable perfect absorber controlled by  the sheet resistance of graphene. Our analytical theory agrees very well with numerical simulations. This approach  is not limited to the studied frequencies and  can be extended to the lower frequencies where kinetic inductance of graphene is very weak. During the talk, we will present more details about the analytical model and  discuss practical implementation of the absorber.

\null\vspace{-7mm}

\begin{acknowledgement}
The authors would like to thank  Mohammad Sajjad Mirmoosa and Ana~D\'{i}az-Rubio for their helpful technical discussions.
This work has received funding from the European Union via the Horizon 2020: Future Emerging Technologies call (FETOPEN-RIA), under grant agreement no. 736876, project VISORSURF.
\end{acknowledgement}

\bibliography{all_angle_perfect_absorption}

\providecommand{\latin}[1]{#1}
\makeatletter
\providecommand{\doi}
  {\begingroup\let\do\@makeother\dospecials
  \catcode`\{=1 \catcode`\}=2 \doi@aux}
\providecommand{\doi@aux}[1]{\endgroup\texttt{#1}}
\makeatother
\providecommand*\mcitethebibliography{\thebibliography}
\csname @ifundefined\endcsname{endmcitethebibliography}
  {\let\endmcitethebibliography\endthebibliography}{}
\begin{mcitethebibliography}{4}
\providecommand*\natexlab[1]{#1}
\providecommand*\mciteSetBstSublistMode[1]{}
\providecommand*\mciteSetBstMaxWidthForm[2]{}
\providecommand*\mciteBstWouldAddEndPuncttrue
  {\def\EndOfBibitem{\unskip.}}
\providecommand*\mciteBstWouldAddEndPunctfalse
  {\let\EndOfBibitem\relax}
\providecommand*\mciteSetBstMidEndSepPunct[3]{}
\providecommand*\mciteSetBstSublistLabelBeginEnd[3]{}
\providecommand*\EndOfBibitem{}
\mciteSetBstSublistMode{f}
\mciteSetBstMaxWidthForm{subitem}{(\alph{mcitesubitemcount})}
\mciteSetBstSublistLabelBeginEnd
  {\mcitemaxwidthsubitemform\space}
  {\relax}
  {\relax}

\bibitem[Wang and Tretyakov(2017)Wang, and Tretyakov]{wang2017tunable}
Wang,~X.-C.; Tretyakov,~S.~A. \emph{arXiv preprint arXiv:1712.01708}
  \textbf{2017}, \relax
\mciteBstWouldAddEndPunctfalse
\mciteSetBstMidEndSepPunct{\mcitedefaultmidpunct}
{}{\mcitedefaultseppunct}\relax
\EndOfBibitem
\bibitem[Zhirihin \latin{et~al.}(2017)Zhirihin, Simovski, Belov, and
  Glybovski]{zhirihin2017mushroom}
Zhirihin,~D.; Simovski,~C.; Belov,~P.; Glybovski,~S. \emph{IEEE Antennas and
  Wireless Propagation Letters} \textbf{2017}, \emph{16}, 2626--2629\relax
\mciteBstWouldAddEndPuncttrue
\mciteSetBstMidEndSepPunct{\mcitedefaultmidpunct}
{\mcitedefaultendpunct}{\mcitedefaultseppunct}\relax
\EndOfBibitem
\bibitem[Tretyakov(2003)]{tretyakov2003analytical}
Tretyakov,~S. \emph{Analytical modeling in applied electromagnetics}; Artech
  House, 2003\relax
\mciteBstWouldAddEndPuncttrue
\mciteSetBstMidEndSepPunct{\mcitedefaultmidpunct}
{\mcitedefaultendpunct}{\mcitedefaultseppunct}\relax
\EndOfBibitem
\end{mcitethebibliography}

\end{document}